\documentclass[twocolumn, prl, superscriptaddress]{revtex4-1}
\usepackage{amssymb}
\usepackage{amsmath}
\usepackage{epsfig}
\usepackage{color}
\usepackage{graphics, graphicx}
\usepackage{bbold}
\usepackage{psfrag}
\usepackage{mathcomp}
\usepackage{subfigure}
\usepackage{verbatim}
\usepackage{color}
\usepackage[colorlinks,citecolor=blue]{hyperref}

\begin{document}
\title{Engineering quantum magnetism in  one-dimensional trapped Fermi gases with p-wave interactions  }
\author{Lijun Yang}
\affiliation{Beijing National Laboratory for Condensed Matter Physics, Institute of Physics, Chinese Academy of Sciences, Beijing 100190, China}
\author{Xiwen Guan}
\affiliation{State Key Laboratory of Magnetic Resonance and Atomic and Molecular Physics, Wuhan Institute of Physics and Mathematics, Chinese Academy of Sciences, Wuhan 430071, China}
\affiliation{Department of Theoretical Physics, Research School of Physics and Engineering, Australian National University, Canberra ACT 0200, Australia}
\author{Xiaoling Cui}
\email{xlcui@iphy.ac.cn}
\affiliation{Beijing National Laboratory for Condensed Matter Physics, Institute of Physics, Chinese Academy of Sciences, Beijing 100190, China}
\date{\today}

\begin{abstract}
The highly  controllable ultracold atoms in a one-dimensional (1D) trap  provide a new platform for  the ultimate simulation of quantum magnetism.
In this regard, the Neel-antiferromagnetism and the itinerant ferromagnetism are of central importance and great interest. Here we show that these  magnetic orders  can be achieved in the  strongly interacting  spin-1/2 trapped Fermi gases with additional p-wave interactions.
In this strong coupling limit, the 1D trapped Fermi gas  exhibit  an effective Heisenberg spin XXZ chain in the anisotropic p-wave scattering channels.
For a particular p-wave attraction or repulsion within the same species of fermionic atoms, the system displays  ferromagnetic domains with full spin segregation or the anti-ferromagnetic spin configuration in the ground state.
Such engineered magnetisms are likely to  be  probed in  a quasi-1D trapped Fermi gas of $^{40}$ K atoms with very close s-wave and p-wave Feshbach resonances.
 \end{abstract}

\maketitle

{\it Introduction.} For several decades the quantum magnetism is one of the central research fields  in condensed matter physics \cite{magnetism}.
In recent years, the ultracold atomic gases provide an ideal platform for the exploration of quantum magnetism  owning to the high controllability of the interaction and the geometry \cite{Esslinger, Hulet}.
In particular,  the Neel anti-ferromagnetism (AFM) in lattices and the itinerant ferromagnetism(FM) in continuum  have been attracting great attention  in both theory and experiment with ultracold atoms.
%
However, the long-range Neel AFM requires sufficiently low temperature associated with the super-exchange coupling in lattices, which is beyond the current cooling technique can achieve so far.
In fact, only the short-range AFM correlation was successfully probed recently \cite{Esslinger, Hulet}.
 The exploration of the itinerant FM in the repulsive branch of three-dimensional (3D) trapped  Fermi gases was unsuccessful, where the possible ferromagnetic state was impeded by severe atom losses when the system  approaches to  the resonance regime\cite{Ketterle1,Ketterle2}.
There have been various proposals for realizing itinerant FM in ultracold atoms, for instance, by using the mass-imbalance\cite{Conduit_mass, Cui_mass, Pilati_mass, Pecak_mass, Zinner_mass}, external potentials\cite{Cui2, Santos,Pu, Zhai_impurity}, SU(N) Fermi gases\cite{Cazallila}, spinor bosons\cite{Levinsen2},  
or multi-orbital\cite{Wu} systems. 
Yet to observe the itinerant FM is still quite challenging in realistic experiment.  

\begin{figure}[t]
\includegraphics[width=8.5cm]{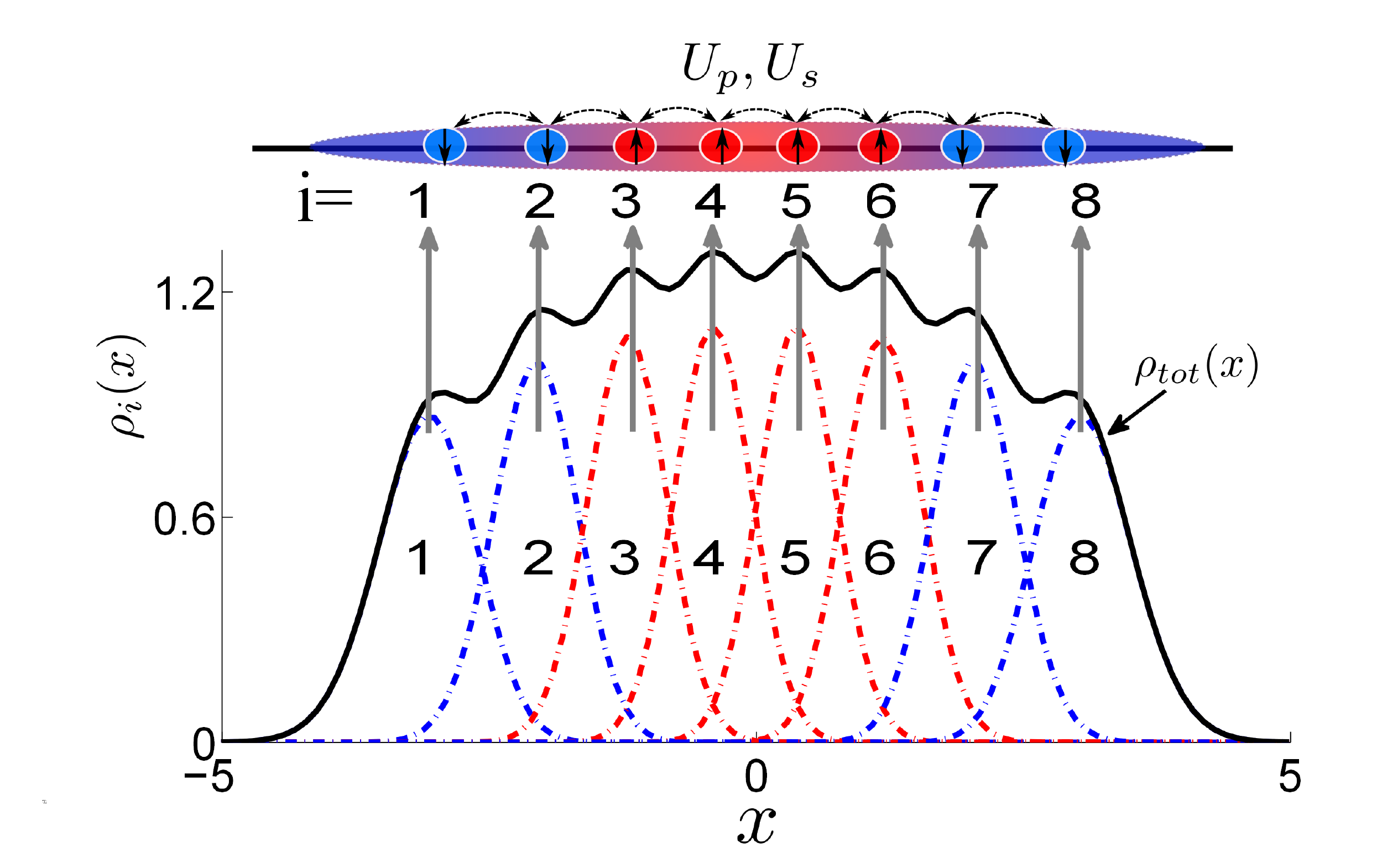}
\caption{(Color online). Density distribution of fermionalized atoms in a harmonic trap. Black solid line is the total density; blue dashed line is for particle at order number $i=1,2,..N (N=8)$ from left to right. $\rho_i$ and $x$ are respectively in the units of $1/a_T$ and $a_T$ ($a_T=1/\sqrt{m\omega_T}$ is the confinement length). Each order number can be effectively mapped to the site index in  a spin chain, with nearest-neighbor coupling determined by the finite s- or p-wave interactions. A weak p-wave attraction between spin-$\uparrow$ atoms will drive the system to itinerant Ferromagnetism with full spin segregation (see the spin distribution along the tube). 
} \label{fig:schematic}
\end{figure}

In this context, the 1D strongly interacting atomic gases offer an alternative while promising opportunity for the simulation of  quantum magnetism in cold atoms, see a recent  review \cite{Guan2013}.
The hidden magnetic spin chain structures for the strongly interacting bosons and fermions
provide insight into the simulation and understanding of FM and AFM orderings \cite{GBT-2007,Furusaki, Guan, Zinner2, Santos, Pu, Levinsen, Zinner3, Levinsen2, Cui1}.
Given the impenetrable feature of particles in a strong coupling limit \cite{Girardeau1,Girardeau2, Sengstock, Chen}, an underlying ``lattice" spin chain can be supported with the site index given by the order  of particles distributed in the spatial space,  see Fig.~\ref{fig:schematic}.
 For a large but finite  s-wave coupling, the 1D trapped Fermi gas is well decoupled into the  charge and spin degrees of freedom with an effective Heisenberg spin chain in the spin sector \cite{Zinner2, Santos, Pu, Levinsen, Zinner3, Levinsen2, Cui1}.
In this effective spin chain, the nearest-neighbor super-exchange interaction  is   fully determined  by the particle collision in the overlap region of their density distributions.
Such a system allows one to  study  quantum magnetism in the absence of lattice potentials\cite{note_NNN}. In particular,  the  atom loss therein can be greatly suppressed by the hard-core interaction\cite{Gora, no_loss},  in contrary to the large atom loss in   a 3D repulsive gas   approaching  to   the resonance \cite{Ketterle2}. 
The effective spin chain can well explain the magnetic correlations in the strongly interacting spin-1/2 trapped fermions \cite{Blume, Cui2, Conduit, Lewenstein, Zinner1}. In these works, the AFM correlation for repulsive fermions \cite{Blume} and the FM transition at the point of full fermionalization\cite{Cui2} were discussed.
Very recently, such a AFM correlation has been experimentally investigated  in a small cluster system of 1D trapped fermions \cite{Jochim_chain}.
    %


In this work, we propose to engineer both the Neel AFM and the itinerant FM  in a strongly s-wave interacting spin-1/2 Fermi gas with additional p-wave interactions in a 1D harmonic trap.
We find a general effective Heisenberg XXZ  model  induced by the anisotropic  p-wave interactions. This model is expected to result in a rich quantum magnetism by adjusting p-wave interaction strengths in different (spin-triplet) channels.
 In particular, we consider the p-wave interaction within one fermion species, as is realizable in the spin-1/2 Fermi gas of $^{40}$K atoms with close s- and p-wave Feshbach resonances \cite{Jin}. We show that by switching on a weak p-wave attraction, the ground state of strongly (s-wave) interacting fermions in the 1D trap  gradually form  the FM domains with full spin segregation; while a weak p-wave repulsion will induce the Neel long-range AFM order.
 %
%
%

{\it The model.} We start from the Hamiltonian for 1D trapped spin-1/2 ($\uparrow,\downarrow$) fermions with both s- and p-wave interactions: $H=H_0+U_s+U_p$,  (we set $\hbar=1$)
\begin{eqnarray}
H_0&=&\sum_i \left( -\frac{1}{2m} \frac{\partial^2}{\partial x_i^2} + \frac{1}{2} m\omega_T^2 x_i^2 \right); \label{H0}\\
U_s&=&  -\frac{2}{ma_{1D}}  \sum_{i,j} \delta(x_{i\uparrow}-x_{j\downarrow}) ; \label{Us} \\
U_p&=& \sum_{i<j} \sum_{M =0,\pm1} -\frac{2l_p^{(M)}}{m} \partial_{x_{ij}}\delta(x_{ij})\partial_{x_{ij}} P_{M}(i,j), \label{Up}
\end{eqnarray}
with $x_{ij}=x_i-x_j$. Here $U_s$ and $U_p$ denote the s- and p-wave interactions with scattering lengths $a_{1D}$ and $l_p^{(M)}$ respectively in the spin singlet and spin triplet (with total magnetization $M=0,\pm 1$) channels, and $P_M$ is the according spin-projection operator in the triplet channel. Note that for  obtaining Eq.3 we have limited ourselves to the weak p-wave interactions \cite{note_lp}.

For the homogenous gas ($\omega_T=0$), the two-body scattering matrix does not comprise the Yang-Baxter equation \cite{Yang1967}.
Thus this  model is not integrable in the presence of both s- and p-wave interactions except few particular cases, i.e. for $l_p^{(M)}=0$ the model reduces to  the Yang-Gaudin model,  whereas for   $a_{1D}=0$ and $l_p^{(0)}=l_p^{(\pm 1)}$ it reduces to the SU(2) spinor Bose gas with a ferromagnetic ground state  \cite{GBT-2007}, also see a recent discussion in \cite{Jiang2015}.
In the following discussion  we will   consider the quasi-1D trapped  Fermi gas with a strong s-wave interaction and a  weak  p-wave interaction, i.e., $|a_{1D}|, |l_p^{M}|$ are  much smaller than the mean particle distance.
The physics in this regime can be well deduced from the fermionalized limit in the framework of perturbation theory, a basic idea for the construction of effective model as following.

{\it The effective spin chain.}  At the point of full fermionization ($a_{1D}=l_p^{(M)}=0$), the ground state of the system is highly degenerate and follows the form of:
\begin{eqnarray}
\Psi_{\xi}&=\phi_F(x_1, x_2, \cdots, x_N)  \langle   \{x_i\}; \{ \mu_i\} |\vec{\xi}\rangle, \label{Psi_f}
\end{eqnarray}
here $\phi_F=\frac{1}{\sqrt{N!}}D(x_1, x_2, \cdots, x_N)$ is the Slater determinant composed by the lowest N-level of eigenstates of $H_0$ and  $\left\{\mu_i \right\}$ denote the spins of the atoms. In the above equations  $|\vec{\xi}\rangle$ is the spin-ordered state 
\begin{eqnarray}
\langle \{x_i\}; \{ \mu_i\} | \vec{\xi} \rangle
=\sum_{P}
 \theta(x_{P_1}< x_{P2}< \cdots <x_{P_N})
\prod_i
 \delta_{\xi_i,\mu_{P_i}},  \label{spin-order}
\end{eqnarray}
describing a sequence of spins  $\xi_{1}$, $\xi_{2},\cdots,\xi_{N}$ placed in order from left to right in the coordinate space. By considering different spin orders $\vec{\xi}$, one can cover all the degenerate ground states. 

In the vicinity of full fermionization (small $a_{1D}$, $l_p^{(M)}$), the particles sitting at the  neighbours   can collide and exchange their spins.
This  process can be well described by an effective spin-chain Hamiltonian by mapping the orders  of particles into site indexes  on a spin-chain.
In the absence of p-wave ($l_p^{(M)}=0$), it is the Heisenberg model associated with s-wave interaction \cite{Zinner2, Santos, Pu, Levinsen, Cui1}, namely  
\begin{equation}
H_{\rm eff}^s= -\frac{ma_{1D}}{2}\sum_lJ_l  \left({\bf s}_l\cdot {\bf s}_{l+1} -\frac{1}{4}\right), \label{Hs}
\end{equation}
where $s_l^{\alpha} (\alpha=x,y,z)$ is the spin operator for atom at site-$l$, and the site-dependent nearest-neighbor coupling is given by
\begin{eqnarray}
J_l&=&\frac{N!}{2} \left(\frac{2}{m}\right)^2 \int d{\bf x}  \Big|\frac{\partial \phi_F}{\partial x_{ij}}|_{x_{ij}=0} \Big|^2 \theta(\cdots<x_i=x_j<\cdots) ,       \label{J_l}
\end{eqnarray}
here $d{\bf x}=\prod_i dx_i$, and $x_i(=x_j)$ is placed in the $l-$th  order in the $\theta-$function.


In the presence of a weak p-wave interaction, we can derive an effective spin-chain Hamiltonian based on the first-order perturbation theory. Given a general ansatz for the many-body wave function
\begin{eqnarray}
\Psi(\{x_i\};\{\mu_i\})=\phi_F(\{x_i\}) \sum_{\xi} a_{\xi} \langle \{x_i\};\{\mu_i\} |\vec{\xi} \rangle,
\label{psi_f_many}
\end{eqnarray}
we can express the energy function produced by p-wave interaction as $E(\{ a_{\xi}\})$. By mapping the spin order number to the site index on a chain, we arrive at  the following effective Hamiltonian which will reproduce the same energy functional $E(\{ a_{\xi} \})$: 
\begin{eqnarray}
H_{\rm eff}^p=\sum_{M=0,\pm 1} \frac{ml_p^{(M)}}{2}\sum_l J_l P_M(l,l+1) . \label{model2}
\end{eqnarray}
where $P_M(l,l+1)$ is the projection operator of the neighboring-site atoms into a spin-triplet with the magnetization $M$ along $z$-direction.
Remarkably, here the coupling $J_l$ follows the same expression as in the s-wave case (Eq.\ref{J_l}), despite of completely different forms between the s- and p-wave contact potentials (Eqs.(\ref{Us},\ref{Up})). The underlying physics for the similar structure of $H_{\rm eff}^s$ and $H_{\rm eff}^p$ share the same spirit as the boson-fermion duality \cite{duality}, i.e.  the bosons with a strong s-wave interaction $g$ is  equivalent to the fermions with a weak p-wave interaction $1/g$.  This mechanism can be generalized to the ``s"-``p" duality in the spin-1/2 fermonic system  in this work. Here  "s" and "p" refer to partial-wave scattering channels.

\begin{widetext}

\begin{figure}[h]
\includegraphics[height=7cm]{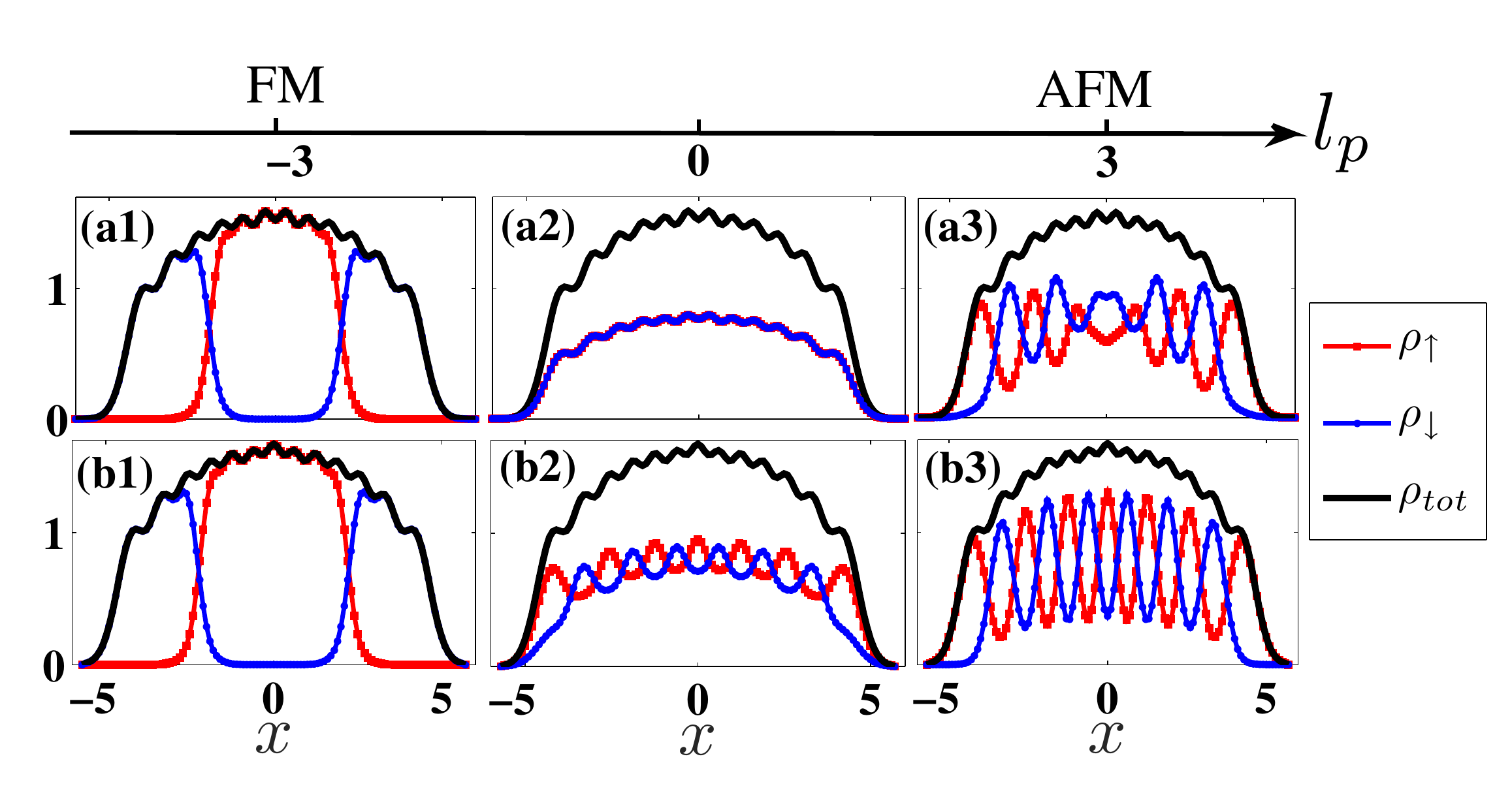}
\caption{(Color online). Magnetic orders induced by p-wave
interaction between spin-${\uparrow}$ atoms. Upper panel (a1-a3) and lower panel (b1-b3) are
respectively the spin density distributions of $6 {\uparrow} +
6{\downarrow}$ and $7{\uparrow} + 6{\downarrow}$ systems. Three
values of $l_p$ (in unit of $-a_{1D}$) are considered: $-3$ (a1,b1), $0$(a2,b2)
and $3$(a3,b3). $\rho_{\sigma}(x)$ and $x$ are respectively in the units of $1/a_{T}$ and $a_T$ ($a_T=1/\sqrt{m\omega_T}$ is the confinement length).  } \label{fig2}
\end{figure}

\end{widetext}

In fact, $H_{\rm eff}^p$ (Eq.\ref{model2}) represents  the Heisenberg spin XXZ model with an effective magnetic field (up to a constant shift):
\begin{equation}
H_{\rm eff}^p=\sum_l j_{l}^{\perp} (s_l^x s_{l+1}^x+s_l^y s_{l+1}^y) +j_l^z s_l^z s_{l+1}^z + h_l (s_l^z + s_{l+1}^z) . \label{model3}
\end{equation}
with $j_{l}^{\perp}=mJ_l  l_p^{(0)}/2,\ j_{l}^{z}=mJ_l(l_p^{(1)}+l_p^{(-1)}-l_p^{(0)})/2\ , h_l=mJ_l (l_p^{(1)}-l_p^{(-1)})/4$.
The fundamental physics of the Heisenberg spin XXZ model can be found in \cite{Takahashi}.
For the isotropic p-wave case with all equal $l_p^{(M)}$, Eq.\ref{model3} reproduces the AFM or FM Heisenberg model for positive or negative $l_p^{(M)}$. For the anisotropic case of $l_p^{(M)}$, which may break both the spin-rotation and the time-reversal symmetry, the system  exhibits a versatile type of magnetic orders depending on the sign and relative strength of $l_p^{(M)}$ in different p-wave scattering channels.

In the rest of the paper, we will consider a realistic case of p-wave interaction in $^{40}K$ Fermi gases, which only exists between one species of fermions (suppose spin ${\uparrow}$)\cite{Jin}. We will show that despite this simple p-wave interaction, the system can display rich and non-trivial magnetic ordering. Our analysis is based on the full effective Hamitonian $H_{\rm eff}=H_{\rm eff}^s+H_{\rm eff}^p$, with $l_p^{(0,-1)}=0$ and $l_p^{(1)}\equiv l_p\neq 0$ in $H_{\rm eff}^p$ (Eq.\ref{model3}). Specifically, we consider a fixed small  negative $a_{1D}$ (strong s-wave repulsion) and a weak tunable $ l_p$ to study how these magnetic orderings occur  in a trapped system with a  given spin number of fermions.


{\it Ferromagnetic vs Antiferromagnetic ordering.} In the p-wave dominated regime, i.e., $|l_p/a_{1D}|\gg 1$, the system can exhibit two types of  magnetic orders depending on the sign of $l_p$:

{\it I. FM for $l_p<0$: } for attractive p-wave interaction, spin $\uparrow$ atoms  tend to line up  in a region with a larger particle (charge) density due to the ferromagnetic ordering 
resulted from  the exchange coupling $J_l$. Thus for a harmonic confinement,  the spin $\uparrow$ atoms  occupy the trap center whereas the  spin $\downarrow$ atoms are repelled to the edges, as being shown by Fig.2 (a1,b1). Here the attractive p-wave interaction within one species of fermions facilitates the formation of Ferromagnetic domains with full spin segregation, i.e., a realization of itinerant FM.

{\it II. Neel AFM for $l_p>0$:}  for repulsive p-wave interaction, spin $\uparrow$ atoms  tend to be separated alternately by spin $\downarrow$ atoms. 
This is a necessary condition for forming the  Neel AFM. 
Indeed, for system with $N_{\uparrow}=N_{\downarrow}+1$, the ground state has the unique Neel AFM spin order $|\{\uparrow\downarrow ...\uparrow\downarrow\uparrow\}\rangle$, as being shown in  Fig.2 (b3). Meanwhile, we note that  
for system with other spin numbers, the ground state of $H_{\rm eff}^p$ could be degenerate.
In particular, for equal mixture $N_{\uparrow}=N_{\downarrow}=N/2$, there is  $(N/2+1)-$fold of degeneracy. For instance, for $N_{\uparrow}=N_{\downarrow}=2$, they are $|\{\uparrow\downarrow\uparrow\downarrow\}\rangle$, $|\{\uparrow\downarrow\downarrow\uparrow \}\rangle$, $|\{\downarrow\uparrow\downarrow\uparrow \}\rangle$. Such degeneracy can be lifted by the presence of $H_{\rm eff}^s$, leading to a unique ground state with the largest weight in $|\{\uparrow\downarrow\downarrow\uparrow \}\rangle$. 
The Neel AFM is  quite prominent in the resulted spin distribution, especially in the outer wings  of the trap, as shown by Fig.2 (a3) for $N_{\uparrow}=N_{\downarrow}=6$. Though we cannot enumerate for all specific combinations of $\{N_{\uparrow}, N_{\downarrow}\}$, we expect the Neel AFM is in general the most dominated magnetic order for repulsive p-wave interaction.


\begin{figure}[t]
\includegraphics[width=9.5cm]{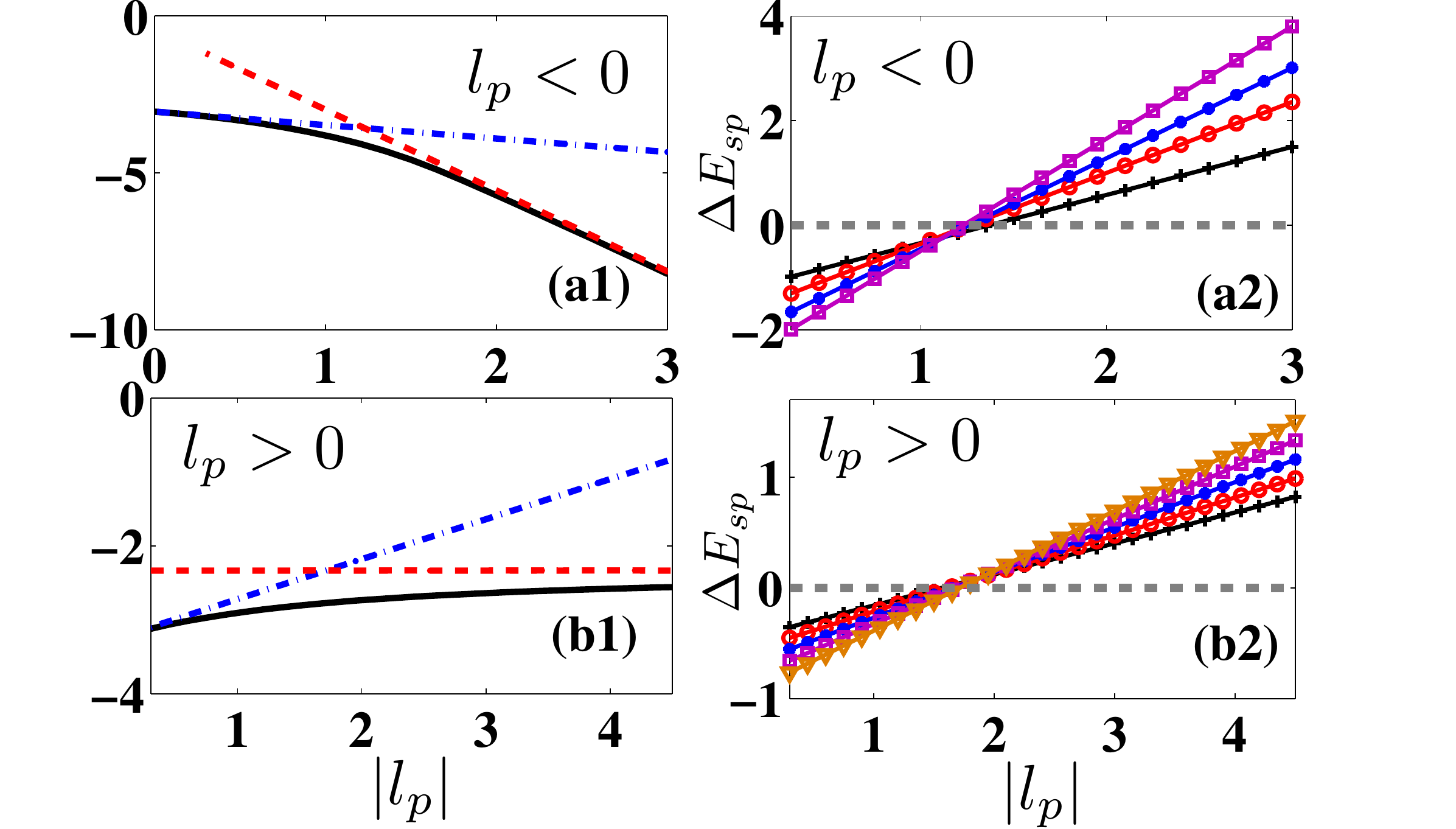}
\caption{(Color online). Crossover from s-wave to p-wave dominated
regime. (a1,b1): The energies $E$ (solid), $E_s$(dash-dot) and $E_p$(dash) as a function of $|l_p|$. (a1) is for $6 {\uparrow} + 6 {\downarrow}$ system with $l_p<0$; (b1) is for $7 {\uparrow} + 6 {\downarrow}$ system with $l_p>0$. 
(a2,b2): $\Delta E_{sp}=E_s-E_p$ as a function of $|l_p|$. In
(a2), $l_p<0$, and the crossing, circle, diamond, and square points are respective for $N_{\uparrow}=N_{\downarrow}=3,4,5,6$ ; in (b2), $l_p>0$, and the crossing, circle, diamond, square and triangle points are respective for $N_{\uparrow}=N_{\downarrow}+1=3,4,5,6,7$. In all plots, the energy unit is
$-ma_{1D}J_1 N/4$, and $|l_p|$ is in unit of $-a_{1D}(>0)$. } \label{fig3}
\end{figure}

In Fig.2, we show the spin density distribution $\rho_{\sigma}(x)=\sum_i n_{i\sigma} \rho_i(x)$, with $n_{i\sigma}$ the density of spin-$\sigma$ at site $i$ obtained from the effective spin-chain Hamiltonian and $ \rho_i(x)$ the particle density at order $i$ in 1D traps\cite{Cui2, Cui1, Santos, Sengstock}. We can see evident spin configuration of FM or Neel AFM being developed as increasing $|l_p/a_{1D}|$ from the s-wave dominated ($l_p=0$ in Fig.2(a2,b2)) to the p-wave dominated(($|l_p/a_{1D}|=3$ in Fig.2 (a1,a3;b1,b3))) regime.

We note that the emergence of above magnetic orders is a crossover rather than a sharp phase transition.
This can be seen  from Fig.3 (a1,b1) that the ground state energy evolves continuously as varying $l_p$.
It is useful to give an estimation to the crossover location, at which point 
the system is expected to enter the p-wave dominated regime and develop an explicit feature of the FM or Neel AFM order in space. In order to do that, we compare two energies: $E_{s/p}=\langle \psi_{s/p} | H_{\rm eff} | \psi_{s/p}\rangle $, where $\psi_s$ and $\psi_p$ are respectively the ground states of $H_{\rm eff}^s$ and $H_{\rm eff}^p$ only, representing the states in the s-wave and p-wave dominated regimes. The crossover value of $l_p$  is then determined by matching $E_s$ with $E_p$. In Fig.3 (a2,b2), we plot the energy difference $\Delta E_{sp}=E_s-E_p$ as a function of $l_p/(-a_{1D})$ for different particle numbers $N$. It is found that the crossover to FM occur at $l_p\approx-1.3|a_{1D}|$ for equal mixtures up to $N=12$, and to AFM occur at $l_p\approx 2|a_{1D}|$ for fermions with $N_{\uparrow}-N_{\downarrow}=1$ up to $N=13$. Therefore these orders can be easily achieved with weak p-wave interactions  given the system is in the strong coupling regime with very small $a_{1D}$.

We remark on several unique advantages for the present scheme to engineer quantum magnetisms in ultracold atoms. 
First, the required conditions, controlled by a single interaction parameter $l_p/a_{1D}$, are practically achievable in current experiments. Second, the resulted ground state is non-degenerate with robust spin configuration (against external perturbations), thanks to the particular p-wave interaction in $^{40}$K and the presence of external trapping potential. Third, one can conveniently achieve distinct magnetic orderings, the Neel AFM and the itinerant FM, in a same system, simply by changing the sign of the p-wave interaction at different sides of the resonance.  Finally, since the weak p-wave interaction will not alter the fermionalized nature of the 1D system, the atom loss is expected to be essentially suppressed  as measured in the hard-core bosons\cite{no_loss}, and thus the stability of the system can be guaranteed.

{\it Experimental relevance.} 
The system described above can be realized using the  $^{40}$K atoms with two hyperfine states $|F=9/2,m_F=-7/2\rangle\equiv|\uparrow\rangle$ and $|F=9/2,m_F=-9/2\rangle\equiv|\downarrow\rangle$. An s-wave Feshbach resonance occurs between $|\uparrow\rangle$ and $|\downarrow\rangle$ at magnetic field $202.1$G (with width 8G), very close to the splitted p-wave resonances at $198.8$G and $198.3$G between two $|\uparrow\rangle$ states\cite{Jin}. Near the p-wave resonances, the s-wave scattering length $a_s\sim 30$nm, which can lead to the strong interaction in quasi-1D geometry through the confinement-induced resonance with transverse confinement length of the same order of $a_s$\cite{Olshanii}. This is achievable by applying deep optical lattices with the confinement length of each harmonic well about one tenth of the lattice spacing $\sim 400$nm. Alternatively, the strong interaction can also be achieved by making the system sufficiently dilute such that $|a_{1D}|$ is much smaller than the inter-particle distance. For the p-wave interaction, as the p-wave Feshbach resonances and its induced 1D resonances \cite{CIR_p_1, CIR_p_2, CIR_p_3} are close to each other and all of them have a resonance width about $0.5$G, a weak p-wave interaction can then be adjusted by fine-tuning the magnetic field around these resonances (with resolution $\lesssim 10$mG\cite{Thywissen}). The predicted FM and Neel AFM order can be probed by imaging the spin density for a many-body system with over hundreds of fermions, or by the tunneling and single-particle level measurements for small cluster systems as in Ref.\cite{Jochim_chain}.

{\it Final remark.} 
Our results shows a powerful realization of  intriguing magnetic orderings in a trapped geometry of ultracold atomic gases with interactions in multiple partial-wave scattering channels. In particular, the inclusion of anisotropic p-wave interactions leads to an effective Heisenberg XXZ spin chain with much richer magnetisms as compared to the pure s-wave interacting case. These results also shed light on the quantum magnetism of fermion systems in higher dimensions. For instance, tuning on a p-wave attraction between one species of fermions will hopefully enhance the ferromagnetic correlation in the repulsive upper branch of a 2D or 3D spin-1/2 Fermi gas, which could be detectable via a pronounced signal of spin fluctuation (as measured in Ref.\cite{Ketterle2}) before the atom losses dominate.


{\bf Acknowledgement.} The work  is supported by the National Natural Science Foundation of China (NNSFC) under grant No.11374177, No. 11421092, No. 11374331 and by the key NNSFC grant No. 11534014, by the National Basic Research Program of China under Grant   No. 2012CB922101 and the programs of Chinese Academy of Sciences.
XWG and XC  thank Y-Z Jiang, D. Kurlov, G. Shlyapnikov and Y.-P. Wang for helpful discussions.


\begin{thebibliography}{99}
\bibitem{note_lp} Such p-wave potential only works for weak interaction giving the Hartree-Fork interaction energy. In the strong coupling regime, the interaction $U_p$ needs to be renormalized, see X. Cui, arxiv: 1605.04363.
\end{thebibliography}
\end{document}